# Design and Analysis of a 170 GHz Antenna for Millimeter-wave Applications

Sheetal Punia, Suman Danani, Hitesh B. Pandya

*ITER-India, Institute for Plasma Research Bhat, Gandhinagar-382428, India*

**ABSTRACT:**

**Microstrip patch antennas are low-profile and robust when mounted on rigid surfaces of the devices making them suitable for communication and millimeter-wave applications. In this paper, an antenna is designed for the resonant frequency of 170 GHz using microstrip technology concerning its miniaturization and cost-effectiveness. The designed antenna is a part of a stray radiation detection system for ECE diagnostic to be installed on fusion research machine ITER. It offers low-directivity to receive radiation from all directions, high bandwidth, low side-lobe-level and return loss of -50 dB, leading to its remarkable utilization in the detection system being designed to protect millimeter wave components of ITER ECE Diagnostic. Power-handling and power-capturing capability of the designed antenna have also been discussed in the paper. Far-field simulations have been performed using CST Microwave Studio software to study the radiation characteristics of the designed antenna.**

## I. INTRODUCTION

Microstrip antennas have various advantages over other conventional antenna techniques due to their simple configuration. They are low-profile (i.e. small height and width), compatible with MMIC designs, quite simple and cheap to manufacture via printed-circuit technology and mechanically robust when placed on rigid surfaces [1]. Furthermore, they also offer some additional properties such as feedline flexibility, dual-polarization, frequency agility and omnidirectional patterning [2,3]. The importance and feasibility of microstrip antenna in our practical world can also be noted by wireless 5G communications and high-performance satellite, medical and missiles applications [4-6]. In this paper, our main goal is to design a compact antenna having wide angle reception. We opted the conventional microstrip technology for the designing purpose as the importance and feasibility of microstrip antenna in our practical world has been proved effectively. However, the designed antenna is based on the conventional Microstrip Patch Antenna design, but its operating frequency stands out the design from the previous work. As we know with an increment in the operating frequency, the dynamics of electrons inside the patch also changes drastically and that can change the whole characteristics of a system. This paper aims to design a reliable antenna array at 170 GHz for stray radiation sensor being designed for ITER ECE Diagnostic and possesses characteristics like low directivity, high bandwidth and better power handling capability that has not been explored till now by any other research group.

The Electron Cyclotron Emission (ECE) diagnostic system on ITER will provide plasma electron temperature profile and electron temperature fluctuations with high spatial and temporal resolution [7-10]. Some challenges need to be addressed to ensure its proper functioning, such as protection from RF stray radiation, obligatory to prevent damage to ECE components [11-13]. These stray radiations are generated due to Electron Cyclotron Resonance Heating (ECRH) wave frequency of 170 GHz utilized for plasma heating and current drive inside the vacuum vessel [14-16]. After multiple–reflections inside the vacuum vessel, the unabsorbed ECRH power becomes diffused and depolarized and may enter the ECE Diagnostic system through port openings. It is proposed to use a sniffer – a real-time power monitoring system- to be followed by a shutter safeguarding the ECE diagnostics. The crucial part of the proposed sniffer is the receiving system that will sense all the power expected to be entering through the port and incident on the ECE components. In this paper, an antenna system is designed to collect the stray electromagnetic energy from the vacuum vessel window and convert it into electrical energy to be monitored by the detector system.

Section II describes the analytical expressions and selection criteria for the characteristic design parameters of the microstrip antenna. These formulas will provide a good estimate of the antenna dimensions intended for millimeter wavelength. The final design of an antenna array is described in Section III. Comparative calculations for the power handling capacity of the individual and array antenna are presented in Section IV. In Section V, simulation results of the antenna are discussed.  Our conclusions are reported in Section VI, highlighting the importance of the designed antenna and scope of future work.

## II. DESIGN PARAMETERS FOR ANTENNA

The microstrip patch antenna is a single-layer design consisting of four main parts: (i) Patch; (ii) Substrate; (iii) ground plane; (iv) feeding part. Interestingly, one can design microstrip antennas for specific resonant frequency, impedance and radiation patterns merely by modifying their dimensions. This feature makes it the best fit for our required application. However, each of these components need to be designed carefully as discussed below:

**Patch Dimensions**

A patch can be of any shape, but the rectangular shape is usually preferred due to its easy fabrication and simple analyses, avoiding complex numerical computations. Considering $\lambda_0 = 1.76mm$ as the free-space incident wavelength, the patch length ($L_p$) and its thickness ($t$) should be in the range $\frac{\lambda_0}{3} < L_p < \frac{\lambda_0}{2}$ and $t \ll \lambda_0$, respectively, for better antenna performance.

**Substrate Selection**

In patch antenna design, the substrate is primarily required to offer mechanical support and spacing between the ground plane and patch. The dielectric constant of the substrate material can alter the electrical performance of the antenna and impedance of transmission line. Therefore, it should be chosen in the range $12 > \varepsilon > 2.2$. For efficient radiation generation, low $\varepsilon$ material is preferred as it enhances the fringe fields that account for the radiation. Furthermore, several quantities must be parameterized for a specific design, such as loss tangent, temperature effects, surface wave excitation, dielectric dispersion, weight, elasticity and cost. Substrates with a higher thickness ($h$) and a high dielectric constant result in smaller bandwidths and lower efficiencies due to the possibility of surface-wave excitation. Hence, it is convenient to keep the height of the substrate in the range of $0.01\lambda_0 \leq h \leq 0.05\lambda_0$ for the significant elimination of multiple surface waves. To overcome some of the problems of thin antenna elements (such as low power handling capability) without unduly sacrificing their principal advantages, they can be built on thick substrates. In this paper, Rogers RO3003 having dielectric constant 3.0, loss factor $tan\delta = 0.001$ (on 10GHz) and $h = 0.254mm$ is chosen as the substrate material to meet our requirements. The patch material used in CST MWS Studio is Copper. The considered Rogers RO3003 high frequency laminates are ceramic-filled PTFE composites for use in printed circuit boards in commercial microwave and RF applications. Its operating frequency range is 30-190GHz.

**Boundary Condition**

As we are interested in measuring the radiated power, all the results for antenna parameters are evaluated in the far-field region determined by the inner boundary distance $s \geq \frac{2D^2}{\lambda_0}$ where, $D = 0.6239\ mm$ is the largest antenna dimension. It gives the far-field distance as $\sim 0.44\ mm$, which is exactly the distance used in simulation.

**Analytical Expressions**

For the analysis of the microstrip antenna, the transmission line method is used for better physical understanding. In this method, the antenna structure is treated as two slots separated by a transmission line. For clarity, the rectangular patch parameters are represented in Fig. 1.

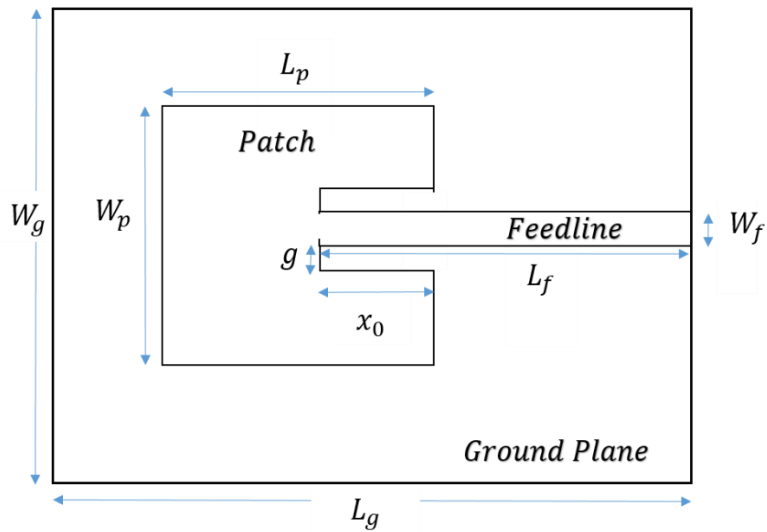

**Figure 1**: Dimensioning sketch for inset fed patch antenna.

Furthermore, the following parameters need to be fixed first while designing an antenna system:

    I.    The dielectric constant of the substrate ($\varepsilon$)
   II.    Resonant frequency ($f$)
  III.    Height of the substrate ($h$)

Once these parameters are fixed, the practical width and length of the patch $W_p$ and $L_p$ are determined by using the following expressions:

$$W_p = \frac{v_0}{2f}\sqrt{\frac{2}{\varepsilon+1}} \tag{1}$$

$$L_p = \frac{1}{2f\sqrt{\varepsilon_{eff}}\sqrt{\mu_0\varepsilon_0}} - 2\Delta L_p \tag{2}$$

where $\Delta L_p$ is the extended length, originated due to fringing effect (i.e., the field lines at the edges of patch undergo fringing) and having dependence on the resonant frequency, as

$$\frac{\Delta L_p}{h} = 0.412\frac{(\varepsilon_{eff}+0.3)\left(\frac{W_p}{h}+0.264\right)}{(\varepsilon_{eff}-0.258)\left(\frac{W_p}{h}+0.8\right)} \tag{3}$$

and, $\varepsilon_{eff}$ is the effective dielectric constant, which accounts for fringing and wave propagation in the feedline defined as

$$\varepsilon_{eff} = \frac{\varepsilon+1}{2} + \frac{\varepsilon-1}{2}\left[1+12\frac{h}{W_p}\right]^{-1/2} \tag{4}$$

Most of the fringing field lines reside in the substrate such that $\varepsilon_{eff} \approx \varepsilon$. The length and width of the ground plane is determined by using the relation $L_g = 6h + L_p$ and $W_g = 6h + W_p$, respectively.

**Microstrip Feedline Dimension**

While designing an antenna, it is necessary to carefully configure the feeding lines in order to prevent impedance mismatching. In other words, antenna parameters need to be optimized by selecting the proper feeding technique and their feed location. The microstrip-line feed technique offers better reliability, ease of fabrication and good polarization purity over other techniques such as coaxial, aperture-coupled, proximity-coupled feed etc. It is etched on the same substrate as that of the patch, providing a planar structure. The width of this microstrip feedline ($W_f$) is related to its characteristic impedance such that

$$Z_T = \frac{60}{\sqrt{\varepsilon_{eff}}} \ln\left(\frac{8h}{W_f} + \frac{W_f}{4h}\right) \tag{5}$$

Microstrip feedline acts as a dispersive transmission line above a critical frequency $f_c \geq 0.3\sqrt{\frac{Z_T}{h}\left(\frac{1}{\sqrt{\varepsilon-1}}\right)}$, where $h$ is in cm. The dispersion in the dielectric constant is given by $\varepsilon_e(f) = \varepsilon - \frac{\varepsilon - \varepsilon_{eff}}{1 + G\left(\frac{f}{f_p}\right)^2}$ where $f_p = \frac{Z_0}{8\pi h}$ and $G = 0.6 + 0.009 Z_0$. $Z_0$ is the static (i.e. low-frequency) characteristic impedance. The phase velocity of quasi TEM-mode propagating in microstrip gets modified as $v_p = \frac{c}{\sqrt{\varepsilon_e(f)}}$ and so does the wavelength $\lambda_p = \frac{v_p}{f}$.

For our antenna design, inset-fed microstrip feed is opted, which has advantages of overcoming narrow-bandwidth constraints and capturing more signals with improved accuracy. In the inset feeding technique, it is preferred to have the feeding location at a certain distance $x_0$ from the edge of the patch towards the center, in order to overcome resonant input resistance. The feeding point can be calculated using the relation

$$R_{in}(x = x_0) = \frac{1}{2G_r} \cos^2\left(\frac{\pi x_0}{L_p}\right) \tag{6}$$

where $G_r$ is the conductance given as $G_r = \frac{1}{90}\left(\frac{W_p}{\lambda_0}\right)^2$.

Using above Eqs. (1) − (6), the dimensions of the inset fed patch antenna are evaluated and tabulated in Table 1.

**Table 1**

Design parameters evaluated for a single patch element.

| S. No. | Parameter | Value |
|---|---|---|
| 1. | Operating frequency $(f)$ | 170 GHz |
| 2. | Thickness $(h)$ | $0.254\ mm$ |
| 3. | Length of Patch $(L_P)$ | $0.3\ mm$ |
| 4. | Width of Patch $(W_p)$ | $0.6239\ mm$ |
| 5. | Cut Width $(g)$ | $0.1\ mm$ |
| 6. | Cut Depth $(x_0)$ | $0.101\ mm$ |
| 7. | Feed Length ($L_f$) | $0.355\ mm$ |
| 8. | Feed Width ($W_f$) | $0.063\ mm$ |

## III.   ANTENNA ARRAY DESIGN

The main characteristics required for our system are wide-angle reception and better power handling capability. Conventional methods used for large apertures are bulky and complicate the design unnecessarily. Antenna array technology entails $N$ number of small-aperture identical antenna elements, exclusively positioned according to their functionality. Therefore, the array provides a large aperture, which makes it proficient to detect extremely weak signals from distant sources with improved gain and radiation pattern compared to a single antenna patch, as shown in Fig. 2. The antenna array also offers better power handling capabilities than a single antenna element that will be discussed in the next section. The feeding location of an antenna array is a crucial part that controls the distribution of voltages among the elements [17]. It is also responsible for proper power distribution among antenna elements due to its steering ability leading to beam phase change [18-20].

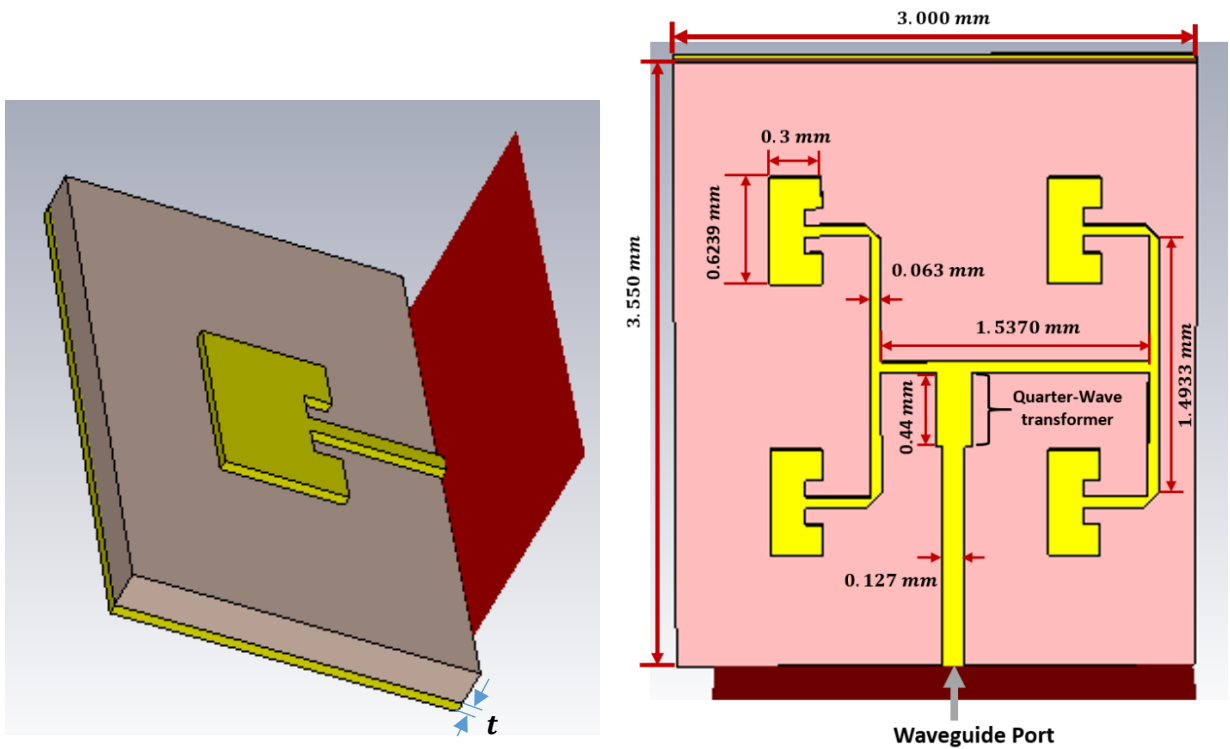

(a) (b)

**Figure 2:** CST model of (a) individual patch and (b) $2 \times 2$ patch antenna array for 170 GHz resonant frequency where $t$ depicts the thickness of the conducting patch.

Depending on the required specifications, one can define various array configurations, such as high bandwidth, high gain and improved efficiency. We designed a $2 \times 2$ patch antenna array for the resonant frequency of 170 GHz, whose dimensions are optimized using Eqs. (1) − (6). Figure 2(b) shows an antenna array displaying measurements of the patch and feeding network etched on the RO3003 substrate with dimensions $3.55 \times 3.0 \times 0.254 \ mm^3$. The substrate thickness is kept small to get higher bandwidth. A symmetric corporate feeding network is used to feed different patches of the antenna array with $50 \ ohm$ microstrip lines of width $0.063 mm$. T- junction technique is adapted for uniform power distribution and compact designing as it reduces the number of power dividers. The distance between two patches should always lie between $\lambda_0/2$ to $\lambda_0$ enabling maximum scanning range and is optimized as $1.6 \ mm$ in this case. The quarter-wave transformer is used to match impedance between the feeding line and the patch to reduce reflection losses, which shall be discussed later. Moreover, the excitation of the system is done by a waveguide port.

## IV.   POWER CALCULATION

**Power Capturing**

The maximum effective aperture of an antenna is defined as $A_{em} = \frac{\lambda_0^2}{4\pi} G_{0r}$ where, $G_{0r}$ is the receiving antenna gain. Using the Friis-Transmission equation, the power collected by the receiving antenna is calculated using the following expression

$$\frac{P_r}{P_t} = \frac{A_{em} G_{0t}}{4\pi R^2} = \left(\frac{\lambda_0}{4\pi R}\right)^2 G_{0r} G_{0t} \qquad (7)$$

where $P_t$ is the input power of the source and $R$ is the distance between source and antenna. For isotropic sources, $G_{0t} = 1$. And $\left(\frac{\lambda_0}{4\pi R}\right)^2$ is a free-space loss factor that considers the losses due to the spherical spreading of the energy by the antenna. Equation (7) relates the power collection capability of an antenna directly to its gain. As we have mentioned earlier, the wide aperture of an antenna array endorses higher gain and hence, higher power collection than a single antenna element.

**Average Power Handling**

The rise in temperature, (i.e., the difference between the maximum operating temperature for substrate ($T_{max}$) and ambient temperature ($T_{amb}$)) may determine the average power handling capability (APHC) of a system as

$$P_{av} = (T_{max} - T_{amb})/\Delta T \tag{8}$$

Here, temperature gradient ($\Delta T$) is the total density of heat flow, having units $^0C/W$, associated with conductor ($\alpha_c$) and dielectric ($\alpha_d$) losses, which are defined as follows

$$\alpha_c = \frac{6.1 \times 10^{-5} Z_0 \epsilon_e \sqrt{\pi f \mu_0/\sigma}}{h} \left[ \frac{W'}{h} + \frac{0.0667 \frac{W'}{h}}{\frac{W'}{h} + 1.444} \right] \left[ 0.5 + \frac{h}{W_p} \left( 1 + \frac{1.25}{\pi} \ln \left( \frac{2h}{t} \right) \right) \right] \tag{9}$$

$$\alpha_d = 27.3 \frac{\epsilon}{\epsilon - 1} \frac{\epsilon_e(f) - 1}{\sqrt{\epsilon_e(f)}} \frac{\tan \delta}{\lambda_0} \tag{10}$$

$$\Delta T = \frac{0.2303 h}{K} \left( \frac{\alpha_c}{W_e(0)} + \frac{\alpha_d}{2 W_e(f)} \right) (^0C/W) \tag{11}$$

The conductivity of the strip conductor is taken as $\sigma = 5.8 \times 10^4 S/mm$ whereas the thermal conductivity ($K$) of dielectric RO3003 is $0.5 W/K/m$. $W_e(0)$ and $W'$ are static and effective width and can be evaluated using $W_e(0) = W_p + \frac{t}{\pi} \left( 1 + \ln \left( \frac{2h}{t} \right) \right)$ and $W' = W_p + 1.25 \frac{t}{\pi} \left( 1 + \ln \left( \frac{2h}{t} \right) \right)$ respectively, where $t = 0.035 mm$ is the thickness of the conducting patch (as shown in Figure 2(a)). The frequency-dependent effective width can be calculated using the expression $W_e(f) = W_p + \frac{W_e(0) - W_p}{1 + \left( \frac{f}{f_p} \right)^2}$. Using above Eqs. (8) – (11), the APHC of individual antenna patch is calculated as $85.6 W$ for the antenna dimensions derived in Section II. For calculations, the ambient temperature is kept as $25^0 C$ whereas the operating temperature is $150^0 C$. Simulations performed for the individual antenna in CST Microwave studio give APHC value 79.5W, closer to the analytical one. In Section III later, it will be seen that the APHC value for the antenna array is $0.76 kW$, which manifests the viability of the antenna array over the individual patch antenna.

## V. RESULTS

Impedance matching always ensures the maximum power transfer between the source and the load. Hence, a noble matching circuit is required for a specific design ensuring its better performance. The corporate feed network can use either a quarter-wave transformer or a stub to match patch element impedance to standard 50-ohm input impedance. The preferences mentioned above for our design are configured in Fig. 3. The important properties of the designed system with both configurations are simulated using CST Microwave Studio software and examined below to check their reliability.

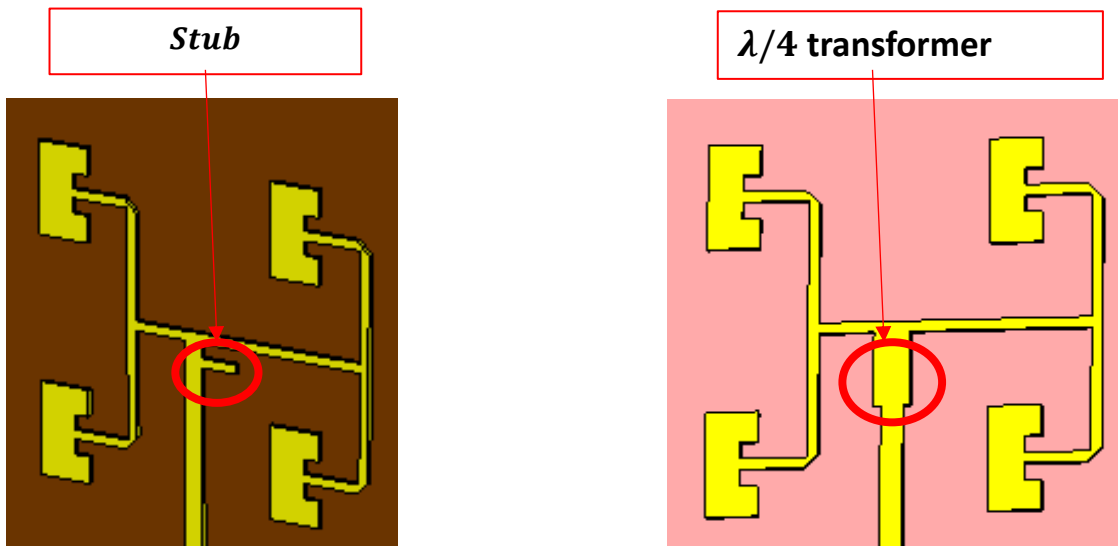

**Figure 3:** Matching configuration for $2 \times 2$ antenna array.

### A. S-Parameter and VSWR

The amount of power reflected from the antenna is parametrized by $S_{11}$ parameter and Voltage Standing Wave Ratio (VSWR) related to each other by the following relation

$$S_{11} = Return\ Loss(dB) = -20 \log_{10} \left(\frac{VSWR - 1}{VSWR + 1}\right) \tag{12}$$

The antenna bandwidth can also be measured from the VSWR plot over a range of frequencies where its value is $\leq 2$. The variation of $S_{11}$ over the range of frequencies for quarter-wave transformer and stub configuration is plotted in Fig. 4. At the resonant frequency, i.e., 170 GHz, the minimum value of $S_{11}$ is $-50\ dB$ for quarter-wave and $-35\ dB$ for stub configuration. It articulates the impedance matching along the propagation path of the signal and effective power delivery for quarter-wave transformer configuration.

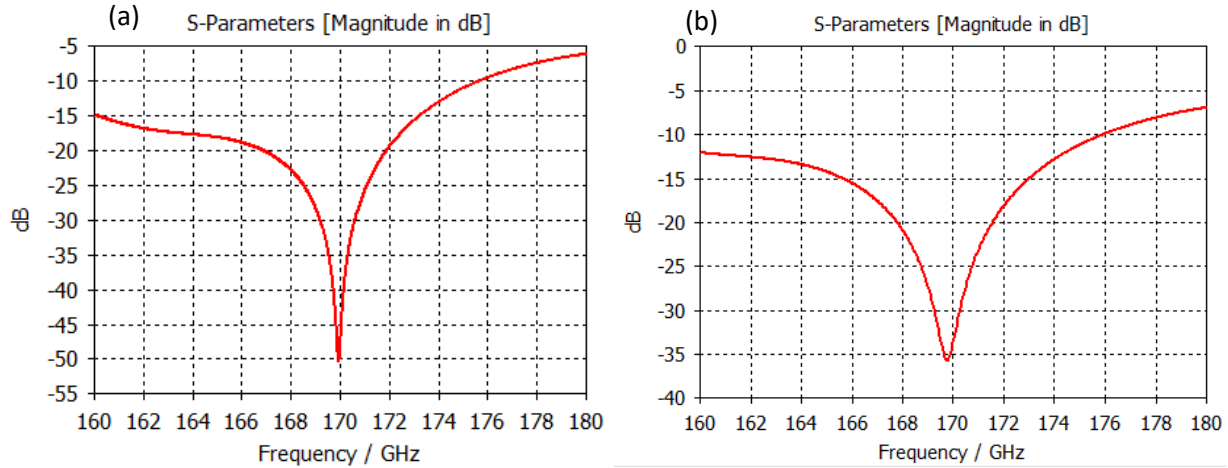

**Figure 4:** Variation of S-parameter with the frequency of incident radiation for (a) Quarter-wave transformer and (b) stub configuration.

### B. Side Lobe Level and Beamwidth

The performance of an antenna is usually measured in terms of gain and its relative 3D radiation pattern. The radiation pattern is represented by polar plots as shown in Fig. 5(a) and (b), measured in the far-field antenna range as mentioned earlier. It suggests that the main lobe direction is different for both the quarter-wave transformer and the stub configuration for the resonant frequency 170 GHz, which is $26^0$ and $7^0$, respectively. However, they possess approximately the same magnitude of gain depicted in Fig. 5(c) and (d). Gain IEEE measured in both configurations does not include the losses associated with the polarization and impedance mismatching.

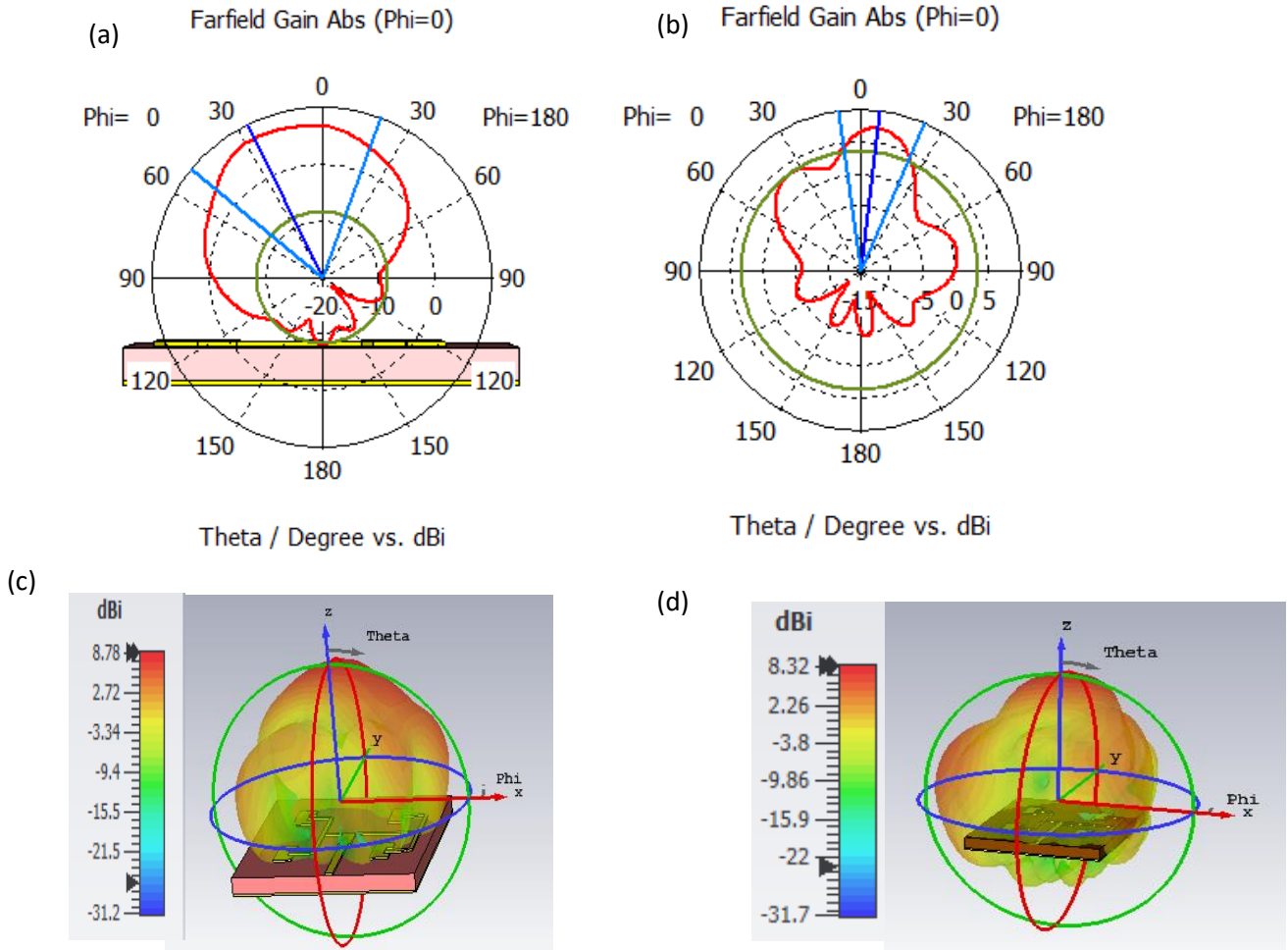

**Figure 5:** Far field characteristics of antenna array for two configurations (i) quarter-wave transformer (a, c); and (ii) stub (b, d).

Far-field characteristics also enable us to investigate two important parameters associated with antenna performance. One is Half-Power beamwidth (also known as angular width), defined as the angular separation of 3dB in the radiation pattern. It deals with the resolution capabilities of the system. And another is the Side-lobe level (SLL), measured by the ratio of the amplitude of the main lobe to that of the side lobe. Low SLL minimizes the false target indication through the side lobes. There is a trade-off between both parameters and one needs to be compromised over the other [21]. We designed a system with a wider reception angle in lieu of high SLL to meet our requirements.

The major difference in both the configuration, i.e., stub and quarter-wave transformer, can be noted in Fig. 5. In the case of stub configuration, the value of angular width and SLL parameter are $31.5^0$ and $-3.6dB$, respectively. However, for quarter-wave transformer these are $70.5^0$ and $-15.5dB$. It is evident from here that the parameters such as bandwidth, gain, S-parameter and side-lobe level attain desirable values for antenna array with quarter-wave transformer as a matching circuit. In other words, the quarter-wave transformer dominates over the stub in terms of better impedance matching, which makes it suitable for our desired application.

### C. Efficiency

The power supplied to the system generally dispenses into surface-wave excitation, radiation emission and conductor and dielectric dissipation. Radiation efficiency is defined as the ratio of power radiated to the power inserted into the antenna system which is plotted in Fig. 6 for the $2 \times 2$ antenna array. It also encompasses the total efficiency of the system that takes into account all $I^2R$ losses and the reflections arise due to mismatching between the antenna and the transmission line.

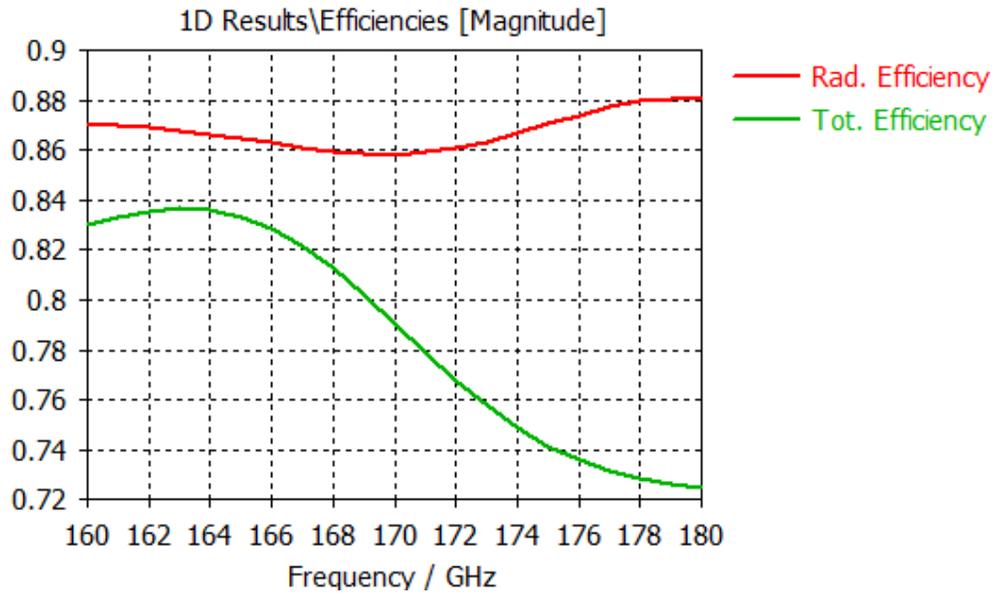

**Figure 6:** Variation of efficiency of antenna array with the frequency.

For better understanding, we summarized the above-discussed properties of the individual patch antenna and the antenna array (with two different matching circuit configurations) in tabular form in Table 2. Here, we conclude the superiority of the 2×2 antenna array with a quarter-wave transformer over other systems in terms of return loss, bandwidth, efficiency, angular width and SLL.

| S. No. | Parameters | $\lambda/4$ Transformer | | Stub |
|---|---|---|---|---|
| | | Antenna Array (h=0.254mm) | Individual Patch (h=0.127mm) | Antenna Array (h=0.254mm) |
| 1. | S-Parameter ($S_{11}$) | -50 dB | -37 dB | -35 dB |
| 2. | VSWR | $\leq 2$ | $\leq 2$ | $\leq 2$ |
| 3. | Bandwidth | 12% | 9 % | 11.2% |
| 4. | Gain (IEEE) | 8.78 dBi | 6.85 dBi | 8.32 dBi |
| 5. | Side Lobe Level (SLL) | -15.5 dB | -12 dB | -3.6 dB |
| 6. | Angular width (3dB) | $70.5^0$ | $77.7^0$ | $31.5^0$ |
| 7. | Total Efficiency (@170GHz) | 79% | 72% | 76% |

**Table 2:** Comparison between individual antenna patch and antenna array with two configurations.

## VI.  CONCLUSION

A $2 \times 2$ microstrip antenna array has been designed and simulated successfully for the resonant frequency of 170 GHz using CST Microwave Studio software. The dominance of the antenna array over the individual patch has been proved while investigating their properties and power handling capabilities. However, continuous increment in the number of patches of an antenna array enhances its directivity further and decreases its angular width. Hence, the $2 \times 2$ antenna geometry reasonably satisfies our desired requirements. Significant properties of the designed system such as S-parameter, VSWR, gain, beamwidth, side-lobe level, efficiency have been estimated and discussed.  A comparative study of the two matching circuits, i.e., stub and quarter-wave transformer, have also been performed. The designed system's power-handling capability is also evaluated to check its reliability. At resonant frequency 170 GHz, the return loss is -50 dB, angular width is $70.5^0$ and the gain is 8.6dBi for the designed antenna. The simulation results show that the designed system fulfills required characteristics and is recommended for high-frequency detector applications.

The future work is to design a detector system for application in ECE diagnostic components protection, utilizing this designed antenna.

-------------------------------------------------